\documentclass{aastex}
\usepackage{emulateapj5}

\newcommand{\bB}{{\bf B}}

\newcommand{\hatk}{{\hat {\bf k}}}

\newcommand{\be}{\begin{equation}}
\newcommand{\ee}{\end{equation}}
\newcommand{\ba}{\begin{eqnarray}}
\newcommand{\ea}{\end{eqnarray}}
\newcommand{\lp}{\left(}
\newcommand{\rp}{\right)}

\newcommand{\etal}{et al.}

\newcommand{\Teff}{T_{\rm eff}}

\newcommand{\Ebp}{E_{Bp}}
\newcommand{\Ebe}{E_{Be}}

\newcommand{\uel}{u_{\rm e}}
\newcommand{\uion}{u_{\rm i}}

\newcommand{\thetakb}{\theta_{kB}}
\newcommand{\taunu}{\tau_\nu^j}
\newcommand{\keso}{\kappa^{\rm es}_0}
\newcommand{\kabs}{\kappa^{\rm a}_j}
\newcommand{\ktot}{\kappa^{\rm t}_j}

\newcommand{\densvp}{\rho_{\rm V}}
\newcommand{\densox}{\rho_{\rm O\rightarrow X}}
\newcommand{\densxo}{\rho_{\rm X\rightarrow O}}
\newcommand{\densx}{\rho_{\rm X,nv}}
\newcommand{\denso}{\rho_{\rm O,nv}}

\newcommand{\kappax}{\kappa_{\rm X}}
\newcommand{\kappao}{\kappa_{\rm O}}
\newcommand{\Ead}{E_{\rm ad}}
\newcommand{\yv}{y_{\rm V}}
\newcommand{\yx}{y_{\rm X}}

\newcommand{\Bpole}{B_{\rm p}}
\newcommand{\Beq}{B_{\rm eq}}
\newcommand{\Tpole}{T_{\rm p}}
\newcommand{\Teq}{T_{\rm eq}}

\newcommand{\holaitwo}{HL03}
\newcommand{\laihoone}{LH02}
\newcommand{\laihotwo}{LH03}

\shorttitle{Spectral Features from Isolated Neutron Stars}
\shortauthors{W.C.G. Ho \& D. Lai}

\begin{document}

\title{Spectral Features in the Thermal Emission from Isolated
 Neutron Stars: Dependence on Magnetic Field Strengths}
\author{Wynn C. G. Ho\altaffilmark{1} and Dong Lai}
\affil{Center for Radiophysics and Space Research, 
Department of Astronomy, Cornell University,
Ithaca, NY 14853}
\email{wynnho, dong@astro.cornell.edu}
\altaffiltext{1}{Current address: Kavli Institute for Particle
Astrophysics and Cosmology,
Stanford University/Stanford Linear Accelerator Center,
P.O. Box 20450, Mail Stop 29, Stanford, CA, 94309}

\begin{abstract}
We study several effects that influence the strength of the proton
cyclotron and atomic features in the thermal spectra of magnetic
neutron stars. We show that
it is possible for vacuum polarization to strongly suppress the
spectral lines
when the magnetic field $B\ga 10^{14}$~G.
For weaker fields ($B\la 7\times 10^{13}$~G),
the surface spectrum is unaffected by vacuum polarization; thus 
the proton cyclotron absorption line can have a large equivalent width.  
Using an approximate calculation of the synthetic spectra, we show that
variation of magnetic fields over the neutron star surface 
leads to broadening of the line. The recently detected
absorption features from the isolated neutron stars
RX~J$1308.6+2127$, RX~J$1605.3+3249$, and RX~J$0720.4-3125$
can plausibly be explained by 
the proton cyclotron resonance, with possible blending due to
atomic transitions, in the atmosphere of the star.
\end{abstract}


\section{Introduction} \label{sec:intro}

In the last few years, considerable observational resources
(e.g., {\it Chandra} and {\it XMM-Newton} telescopes)
have been devoted to the study of thermal emission from isolated
neutron stars (NSs) and, in particular, to the search for
spectral features in the radiation. For many NSs,
the spectra are found to be featureless and often well
fit by a blackbody (e.g., Marshall \& Schulz~2002;
Burwitz \etal~2003; see Pavlov, Zavlin, \& Sanwal~2002 for a review).

Recently, absorption features have been found in the 
thermal emission of several isolated NSs. For example, 
the spectrum of the young NS 1E~$1207.4-5209$ (with 
surface temperature $T_{\rm eff}\simeq 2$~MK)
shows features at 0.7 and 1.4~keV
(Sanwal \etal~2002; Mereghetti \etal~2002; Hailey \& Mori 2002)
and possibly at 2.1 and 2.8 keV (Bignami \etal~2003).
Two NSs, RX~J$1308.6+2127$ (= RBS~1223) 
and RX~J$1605.3+3249$, belonging to the class
of dim, radio-quiet isolated NSs (see Haberl~2003), 
have been observed to possess broad absorption features in their spectra:
at $\simeq 0.2-0.3$~keV for RX~J$1308.6+2127$
(Haberl \etal~2003a) and at $\simeq 0.45$~keV 
for RX~J$1605.3+3249$ (van Kerkwijk \etal~2004).
These are in contrast to two similar dim NSs,
RX~J$1856.5-3754$ (Pons \etal~2002; Burwitz \etal~2003)
and RX~J$0720.4-3125$ (Paerels \etal~2001), which show 
featureless, blackbody-like spectra (however, see Haberl \etal~2003b
for the detection of a line at 0.27~keV in RX~J$0720.4-3125$).
It is particularly striking that, although RX~J$0720.4-3125$,
RX~J$1605.3+3249$, and RX~J$1308.6+2127$ have similar effective
temperatures ($\simeq 1$~MK), 
the equivalent widths (EW) of their lines are very different
(van Kerkwijk~2003),
from weakest in RX~J$0720.4-3125$ (EW $\approx 40$~eV)
to stronger in RX~J$1605.3+3249$ (EW $\approx 80$~eV)
to the strongest in RX~J$1308.6+2127$ (EW $\approx 150$~eV).
Motivated by these observations, we study in this
paper the strength of the proton cyclotron line (and atomic lines) 
in the atmosphere emission of NSs as a function of magnetic field strength. 
We show that the absorption features observed in the three NSs 
(RX~J$1308.6+2127$, RX~J$1605.3+3249$, RX~J$0720.4-3125$) can
possibly be explained
by the proton cyclotron resonance at (unredshifted) energy
\be
\Ebp={\hbar eB\over m_pc}=0.63\,B_{14}~{\rm keV},
\ee
where $B=10^{14}B_{14}$~G is the magnetic field strength.
(There may be blending of absorption features from radiative transitions 
in neutral H atoms). We also show that, owing to the vacuum polarization
effect, the EW of the line can decrease dramatically as the magnetic
field enters the magnetar regime ($B\ga 10^{14}$~G).

In \S\ref{sec:vp}, we review and elaborate upon the effect of vacuum
polarization on radiative transfer in NS atmospheres. 
In \S\ref{sec:spectra}, we present our numerical results for the atmosphere
spectra with different magnetic field strengths; these results illustrate the
suppression of atmosphere line features as the field strength increases. 
We consider synthetic spectra from NS atmospheres in \S\ref{sec:synth} and 
discuss our results in \S\ref{sec:discussion}.

\section{Vacuum Polarization Effect on 
Radiative Transfer in Neutron Star Atmospheres}
\label{sec:vp}

Before presenting our numerical results of NS atmospheric spectra
for different magnetic field strengths (\S\ref{sec:spectra}),
we first review and elaborate
upon the key physics of the vacuum polarization effect on radiative
transfer in strong magnetic fields (see Lai \& Ho~2002, Ho \& Lai~2003,
hereafter \laihoone, \holaitwo;
see especially \S2 of Lai \& Ho 2003a, hereafter \laihotwo).

\subsection{Mode Conversion: Two Magnetic Field Regimes}
\label{sec:mc}

Quantum electrodynamics predicts that 
in strong magnetic fields the vacuum becomes birefringent.
In a magnetized NS atmosphere, both the plasma and vacuum polarizations
contribute to the dielectric property of the medium. A ``vacuum
resonance'' arises when these two contributions ``compensate'' each other
(Gnedin, Pavlov, \& Shibanov~1978; M\'{e}sz\'{a}ros \& Ventura~1979;
Pavlov \& Shibanov~1979; Ventura, Nagel, \& M\'{e}sz\'{a}ros~1979).
For a photon of energy $E$, the vacuum resonance occurs at the density 
\be
\densvp\simeq 0.964\,Y_e^{-1}B_{14}^2E_1^2 f^{-2}~{\rm g~cm}^{-3},
\label{eq:densvp}
\ee
where $Y_e$ is the electron fraction, $E_1=E/(1~{\rm keV})$,
and $f=f(B)$ is a slowly varying function of $B$ and is of order unity
(\laihoone; \holaitwo). For $\rho>\densvp$ (where the plasma effect
dominates the 
dielectric tensor) and $\rho<\densvp$ (where vacuum polarization dominates), 
the photon modes (for $E\ll\Ebe=1.16\,B_{14}$~MeV, the electron cyclotron 
energy) are almost linearly polarized: the extraordinary 
mode (X-mode) has its electric field vector 
perpendicular to the $\hatk$-${\hat\bB}$ plane, while the ordinary mode 
(O-mode) is polarized along the $\hatk$-$\hat\bB$ plane
(where $\hatk$ specifies the direction of photon propagation and $\hat\bB$ 
is the unit vector along the magnetic field).  Near $\rho=\densvp$, however,
the normal modes become circularly polarized as a result of the
``cancellation'' of the plasma and vacuum effects.
When a photon propagates outward in the NS atmosphere, its polarization
state will evolve adiabatically if the density variation is
sufficiently gentle. Thus, a X-mode (O-mode) photon will be converted
into a O-mode (X-mode) as it traverses the vacuum resonance.
For this conversion to be effective, 
the adiabatic condition must be satisfied:
\be
E\ga \Ead =2.55\,\bigl(f\,\tan\thetakb |1-\uion|\bigr)^{2/3}
\left({1\,{\rm cm}\over H_\rho}\right)^{1/3}~{\rm keV},
\label{eq:enad}
\ee
where $\thetakb$ is the angle between ${\hatk}$ and ${\hat\bB}$, $\uion
=(\Ebp/E)^2$, and $H_\rho=|dz/d\ln\rho|$ is the density scale
height (evaluated at $\rho=\densvp$) along the ray.
For an ionized hydrogen atmosphere, $H_\rho\simeq 2kT/(m_pg\cos\theta)
=1.65\,T_6/(g_{14}\cos\theta)$~cm, where $T=10^6\,T_6$~K is the temperature,
$g=10^{14}g_{14}$~cm~s$^{-2}$ is the gravitational acceleration, and $\theta$
is the angle between the ray and the surface normal.
In general, the mode conversion probability is given by 
$P_{\rm con}=1-e^{-(\pi/2)(E/\Ead)^3}$.

\begin{figure*}
\plotone{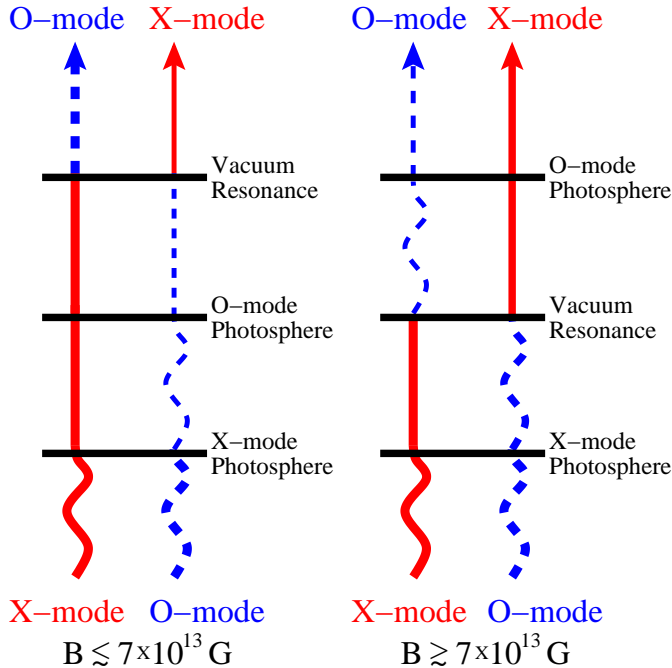}
\caption{
A schematic diagram illustrating how vacuum polarization-induced
mode conversion affects the emergent radiation from a magnetized
NS atmosphere.
The photosphere is defined by where the optical depth (measured from
the surface) is 2/3 and is where the photon decouples from the matter.
The left side applies to the ``normal'' field regime
[$B\la 7\times 10^{13}$~G; see eq.~(\ref{eq:maglim})], in which the 
vacuum resonance lies outside the photospheres of the two modes.
The right side applies to the ``superstrong'' field regime
($B\ga 7\times 10^{13}$~G), in which the vacuum resonance lies
between the two photospheres.
\label{fig:mc}
}
\end{figure*}

Because the two photon modes have very different opacities,
the vacuum polarization-induced mode conversion can significantly affect
radiative transfer in magnetized atmospheres.
When the vacuum polarization effect is neglected (nv),
the decoupling densities of the O-mode and X-mode photons 
(i.e., the densities of their respective photospheres)
are approximately given by (see \laihoone)
\ba
\denso & \approx & 0.42\,T_6^{-1/4}E_1^{3/2}S^{-1/2}\mbox{ g cm$^{-3}$}
\label{eq:denso} \\
\densx & \approx & 486\,T_6^{-1/4}E_1^{1/2}S^{-1/2}B_{14}\mbox{ g cm$^{-3}$},
\label{eq:densx}
\ea
where 
$S=1-e^{-E/kT}$.
Thus the X-mode photons are produced in deeper, hotter layers in the 
atmosphere than the O-mode photons. 
When vacuum polarization is taken into account, 
the decoupling densities can be altered depending on the location
of the vacuum resonance $\densvp$ relative to 
$\denso$ and $\densx$ (see Fig.~\ref{fig:mc}).
For ``normal'' magnetic fields, 
\be
B < B_l \approx 6.6\times 10^{13}\, T_6^{-1/8}E_1^{-1/4}S^{-1/4}
\mbox{ G},\footnote{Since $B_l$ depends weakly on $T$, one may use
$\Teff/(10^6\mbox{ K})$ as an estimate for $T_6$.  They only differ by a
factor of a few in the cases considered here.}
\label{eq:maglim}
\ee
the vacuum resonance lies outside both photospheres ($\densvp<\denso<\densx$);
for the magnetar field regime, $B>B_l$, the vacuum resonance lies between 
these two photospheres ($\denso<\densvp<\densx$). These two field regimes
correspond to qualitatively different vacuum polarization effects on the
radiative transfer.

We can estimate the effective decoupling (photosphere) densities of the two
photon modes when mode conversion at the vacuum resonance is taken into
account (see, e.g., \laihoone). Consider first the ``normal'' field regime
$B<B_l$.  The free-free opacities of the two modes are approximately given by 
(neglecting angle-dependent factors)
$\kappao\simeq 9.3\,\rho_1\,T_6^{-1/2}E_1^{-3}S$
cm$^2$g$^{-1}$
and $\kappax\simeq \kappao \uel^{-1}$, where
$\rho=\rho_1\,\mbox{g cm$^{-3}$}$ and $\uel=(\Ebe/E)^2\gg 1$.
From hydrostatic equilibrium, the column density $y$ at density $\rho$ is
$y=0.83\,\rho_1T_6$~g~cm$^{-2}$ (for $g_{14}=2$).
The column depth of the X-mode photosphere is determined by
$\int_{\yv}^{\yx}\!\!\kappax dy+\int_0^{\yv}\!\!\kappao dy=2/3$.
Thus the photosphere density of the X-mode photons (which 
convert to the O-mode upon traversing the vacuum resonance) is 
\ba
\densxo&\simeq & \left[\densx^2+\densvp(1-\uel)\right]^{1/2}
\nonumber\\
&\simeq & \densx\left[1-(B/B_l)^4\right]^{1/2}.
\ea
Similarly, the photosphere density of the O-mode photons (which convert
to the X-mode at the resonance) is
\be
\densox
\simeq \denso\left[1+(B/B_l)^4\right]^{1/2}.
\ee
Thus, when $(B/B_l)^4\ll 1$, the photospheres of the two modes are
given by equations~(\ref{eq:denso}) and
(\ref{eq:densx}), and there is no net change in
the total emission spectrum. We note, however, that the X-ray 
polarization signals are dramatically affected by vacuum polarization
in this ``normal'' field regime (see Lai \& Ho 2003b).

For the magnetar field regime, we find that the effective photosphere
densities are given by
\ba
\densox & \simeq & \densvp\left[1+(B_l/B)^4\right]^{1/2} \label{eqox} \\
\densxo & \simeq & \denso.
\ea
Thus for $(B/B_l)^4\gg 1$, the O-mode photosphere is unchanged
by vacuum polarization, while the X-mode photons (which carry the bulk of the
flux) emerge from the vacuum resonance layers.
We expect that the total spectrum will be affected
by vacuum polarization in this regime (see \S\ref{sec:spectra}).

\subsection{No Mode Conversion}
\label{sec:nc}

The numerical results shown in \S\ref{sec:spectra}
are based on the ``full mode conversion'' ansatz, i.e., 
when we treat the vacuum polarization effect, 
mode conversion is assumed for all photon energies and directions of
propagation. In reality, mode conversion is complete only when 
the adiabatic condition is satisfied [see eq.~(\ref{eq:enad})].
A proper treatment of partial mode conversion in radiative
transfer requires solving the Stokes parameters 
of the radiation field, rather than the intensities of the two
photon modes (see \laihotwo).  The current generation of NS 
atmosphere models cannot properly treat partial mode
conversion.  We note that, in the opposite limit
(i.e., the complete non-adiabatic limit or when mode conversion at the
vacuum resonance is turned off), the X-mode photon
decoupling depth can still be affected by vacuum polarization.
This is because the X-mode opacity (without mode conversion) exhibits a
sharp rise near the resonance. As a result, the X-mode decoupling density
is close to $\densvp$ for sufficiently high photon energies
(see \laihoone\ and \holaitwo).
As discussed in \S4 of \holaitwo, the proton cyclotron line can be
suppressed even in the non-adiabatic limit,
when the magnetic field is in the range $\mbox{a few}\times 10^{14}
\mbox{ G}\la Bf^{-4/5}T_6^{1/10}\la 2\times 10^{15}$~G.  (This range
of magnetic fields is very approximate since they depend on the direction
of photon propagation and redistribution of photon spectral flux occurs
in real atmospheres.)

By assuming full mode conversion in \S\ref{sec:spectra},
we are artificially exaggerating the vacuum polarization effect on the
emergent spectrum; despite this, we expect the trend (i.e., the
magnetic field dependence of the spectral features) described in
\S\ref{sec:spectra} to be correct (see \holaitwo\ for a comparison of 
numerical results based on the ``full mode conversion'' ansatz and the ``no
conversion'' ansatz).  Most importantly, near the ion cyclotron energy
($\uion\approx 1$), the adiabatic condition is always satisfied
[see eq.~(\ref{eq:enad}); see especially Fig.~6 of \laihotwo].
For example, at $B=7\times 10^{13}$~G ($\Ebp=0.44$~keV),
the adiabatic condition is satisfied when
$E>0.40$~keV and $E>0.42$~keV for
$\thetakb=15^\circ$ and $30^\circ$, respectively;
at $B=10^{14}$~G ($\Ebp=0.63$~keV),
eq.~(\ref{eq:enad}) is satisfied when
$E>0.54$~keV and $E>0.58$~keV for
$\thetakb=15^\circ$ and $30^\circ$, respectively.
Thus we expect the mode conversion calculation to be more accurate.

\section{Atmosphere Spectra: Dependence on Magnetic Field Strength}
\label{sec:spectra}

\begin{figure*}
\plotone{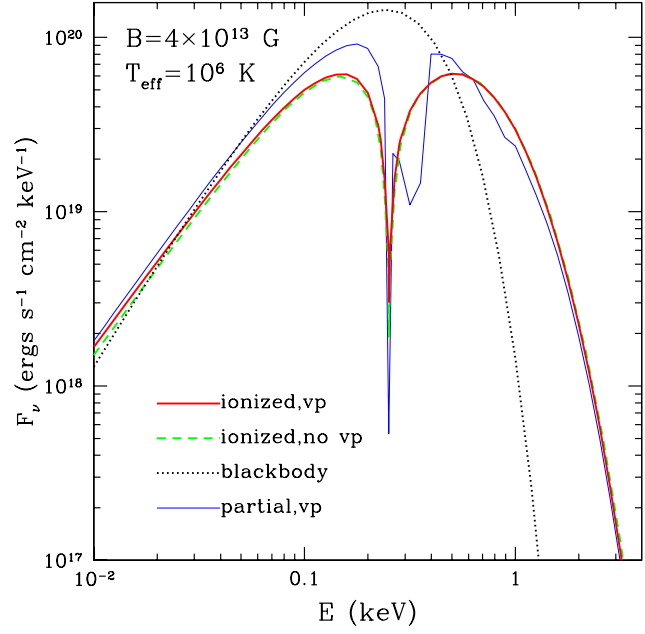}
\caption{
Spectra of fully ionized hydrogen atmospheres with
$\Teff=10^6$~K and $B=4\times 10^{13}$~G.
The thick solid line refers to the atmosphere with the 
vacuum polarization effect included (vp),
the dashed line refers to the atmosphere without vacuum
polarization (no vp), and the dotted line is for a blackbody with $T=10^6$~K.
The result for a partially ionized
atmosphere including vacuum polarization is also shown (thin solid line).
\label{fig:spectra}
}
\end{figure*}

\begin{figure*}
\plotone{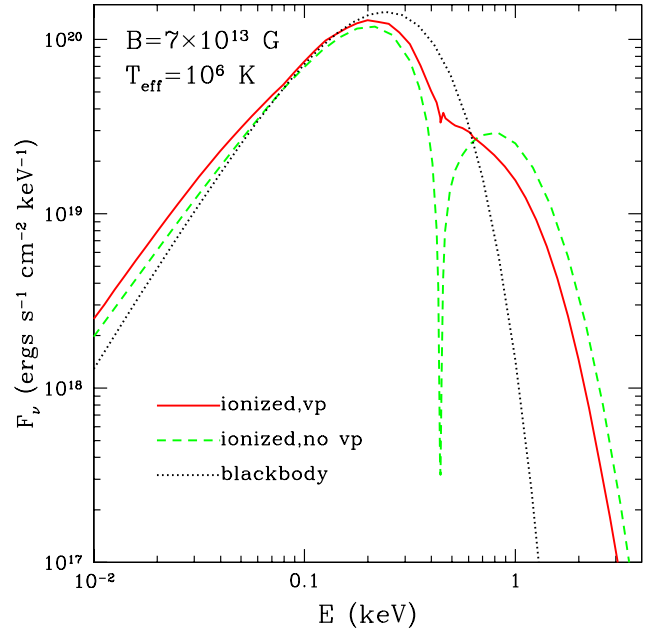}
\caption{Same as Fig.~\ref{fig:spectra}, except for $B=7\times 10^{13}$~G.
\label{fig:spectra2}
}
\end{figure*}

\begin{figure*}
\plotone{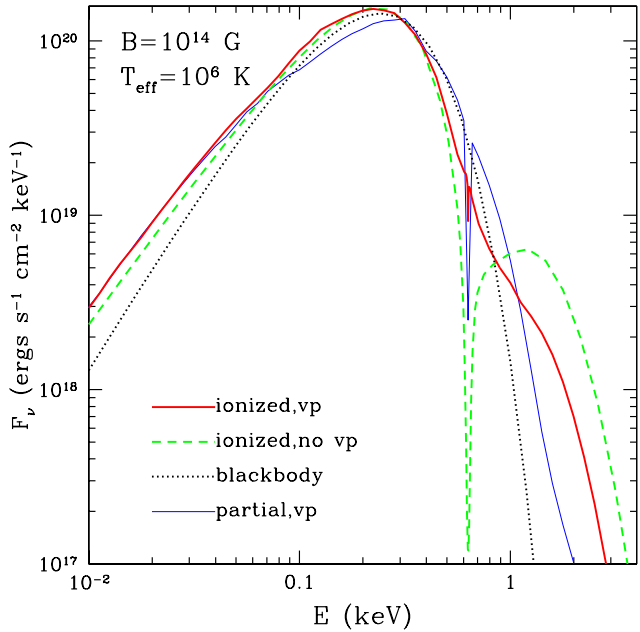}
\caption{Same as Fig.~\ref{fig:spectra}, except from $B=10^{14}$~G.
\label{fig:spectra3}
}
\end{figure*}

Figures~\ref{fig:spectra}-\ref{fig:spectra3} depict the spectra of
ionized hydrogen atmospheres
for different field strengths, all with $T_{\rm eff}=10^6$~K
and the magnetic field perpendicular to the stellar surface.
These correspond to emission from a local patch on the NS surface
and are calculated using atmosphere models
in radiative equilibrium (see Ho \& Lai~2001 and \holaitwo\ for details).
When the vacuum polarization effect is neglected (see the curves 
labeled ``no vp''), all spectra exhibit a broad absorption 
feature due to the proton cyclotron resonance (Ho \& Lai~2001; 
cf. Zane \etal~2001; Lloyd~2003). When the vacuum polarization effect is
included, the spectra for different field strengths can be very different.
For $B=4\times 10^{13}$~G, vacuum polarization produces
no measurable difference in the spectrum (see Fig.~\ref{fig:spectra}),
although we note that the vacuum effect increases the temperature
of the atmosphere layers outside the photospheres of both photon modes
(see \holaitwo).
For $B=10^{14}$~G (see Fig.~\ref{fig:spectra3}),
which satisfies $(B/B_l)^4\gg 1$,
the effective X-mode photosphere is located at the vacuum resonance density
where mode conversion takes place [$\densox\simeq
\densvp$; see eq.~(\ref{eqox})]. Since the X-mode photons (which carry the
bulk of the thermal energy) emerge from shallower layers of the 
atmosphere (compared to the no vacuum polarization case),
the high-energy spectral tail is softened, and the spectrum is
closer to a blackbody (although the spectrum
is still harder than blackbody because of non-grey opacities).
Also, for magnetar field strengths, 
since photons with energies around $\Ebp$ (both near the cyclotron 
line center and outside the line) decouple
from similar depths (i.e. at density $\densvp$) in the atmosphere,
the proton cyclotron line is suppressed by the 
vacuum polarization effect
(cf. Zane \etal~2001; Lloyd~2003).
We see from Fig.~\ref{fig:spectra2} that the $B=7\times 10^{13}$~G case is
intermediate between the ``normal'' field regime and the magnetar field regime.


For the temperature and magnetic field strengths considered in
Figs.~\ref{fig:spectra}-\ref{fig:spectra3}, 
the fraction of neutral H atoms in the atmosphere is not negligible,
and atomic bound-bound and bound-free transitions may also give rise
to absorption features. However, partially ionized magnetic atmosphere models
are not as well developed as fully ionized models (see Ho \etal~2003).
In Figs.~\ref{fig:spectra} and \ref{fig:spectra3}, we also show the spectra of
partially ionized H models 
for $B=4\times 10^{13}$~G and $B=10^{14}$~G, including
the effect of vacuum polarization (using the ``full mode conversion'' ansatz); 
these are based on the latest opacity and equation of state results
for magnetized H plasmas (Potekhin \& Chabrier 2003, 2004).  For
$B=4\times 10^{13}$~G, the absorption feature
associated with the ground-state bound-free transition lies at
$\simeq 0.44$~keV, and the bound-bound transition
(from $s=0,\nu=0$ to $s=1,\nu=0$) is at 0.35~keV, which is
close to the proton cyclotron resonance; these features are unaffected
by the vacuum polarization effect.  For $B=10^{14}$~G, the 
bound-free feature is at $~0.54$~keV, and the bound-bound transition
is at 0.74~keV, which is also close to $\Ebp$.
Like the proton cyclotron resonance in the magnetar field regime,
these atomic features are also suppressed by the vacuum polarization
effect.

The results shown in
Figs.~\ref{fig:spectra}-\ref{fig:spectra3} assume the adiabatic
condition is always satisfied [see eq. (\ref{eq:enad}) and \S\ref{sec:nc}].
Numerical models, with $B\le 10^{14}$~G and $\Teff=10^6$~K
and assuming no mode conversion, do not show suppression of the
proton cyclotron line since there is negligible photon flux at
high energies to be redistributed to lower energies;
line suppression (with no mode conversion) starts becoming important at
$\Teff\approx 3\times 10^6$~K for $B\approx 10^{14}$~G
(see, e.g., Fig.~8 of Ho \etal~2003).
Nevertheless, as discussed in \S\ref{sec:nc}, we believe the full mode
conversion calculation to be more accurate (since the adiabatic condition
is always satisfied) near the proton cyclotron energy, and the
vacuum polarization-induced line suppression shown here should occur.

\section{Synthetic Spectra from Neutron Star Surfaces} \label{sec:synth}

The spectra presented in \S\ref{sec:spectra} correspond to emission
from a local patch of the NS surface. The proton cyclotron line in the local 
spectrum exhibits a rather spiky feature, which is clearly not 
observed (see also Ho \& Lai 2001; Zane \etal~2001; Lloyd~2003).
To compare with observations, 
we need to calculate the synthetic spectrum from the whole NS surface.
Such synthetic spectra is necessarily model-dependent
(see Zane \etal~2001 for a different model and method),
as the variations of magnetic field and temperature on the NS surface
are unknown.

Here, for illustrative purposes, we assume a dipole magnetic
field configuration and consider two models for the 
temperature variation: (1) a uniform surface 
temperature over the entire NS surface $T_0(\tau)$;
(2) a (somewhat arbitrarily-chosen) non-uniform temperature
distribution given by
\be
T(\tau,\gamma) = \Tpole(\tau)\cos^2\gamma + \Teq(\tau)\sin^2\gamma
 \label{eq:tempdist},
\ee
where $\gamma$ is the magnetic colatitude, $\tau=-\int\rho\keso\,dr$
is the Thomson depth in the atmosphere, and $\keso$
is the Thomson opacity.
We generate the temperature profiles $\Tpole(\tau)$
and $\Teq(\tau)$ from our atmosphere models for
effective temperatures $\Teff$ and magnetic fields $B$
given by $(\Teff,B)=(\Teff^{\rm p},\Bpole)$ and
$(\Teff^{\rm eq},\Beq)$, respectively, 
and we assume $T_0(\tau)$ is the same as $\Tpole(\tau)$. 

Let the angle between the line of sight toward the observer (the $Z$-axis)
and the axis of the magnetic pole be $\Theta$. (For a spinning  NS, $\Theta$ 
obviously varies as the star rotates.) The observed spectral 
flux can be calculated using the standard procedure, including
general relativistic effects (Pechenick, Ftaclas, \& Cohen~1983;
see also Pavlov \& Zavlin 2000). At a given local patch on the NS surface, 
we denote the direction of the photons which enter the detector
by $\hatk$; the angle between $\hatk$ and the
local surface normal is $\theta$ and the azimuthal angle of the emitting
patch is $\phi$. 
For a given mode $j$, we approximate the specific
intensity of the radiation from this patch by 
$B_\nu(T)/2$ (neglecting the correction to the source function due to
scattering), where $B_\nu(T)$ is the Planck function and
$T$ is the temperature at $\taunu(\theta,\phi)=2/3$. Here, 
$\taunu(\theta,\phi)$ is the effective optical depth along the ray,
\be
\taunu(\theta,\phi) = \frac{1}{\keso}
\int_0^\tau\lp\kabs\ktot\rp^{1/2}\frac{d\tau'}{\cos\theta}, 
\ee
where $\kabs$ and $\ktot$ are the absorption and total opacities,
respectively. Interpolating from a table of calculated optical depths
$\taunu(\tau,\theta,\phi)$, we determine the value of $\tau$ at which 
$\taunu=2/3$, thereby obtaining the source function
for each mode. The observed flux spectrum $f_{\nu_0}$ 
(at the observed frequency $\nu_0$) from the entire NS surface
is then $f_{\nu_0}=f^{\rm X}_{\nu_0}+f^{\rm O}_{\nu_0}$, with
\be
f^j_{\nu_0} =(1+z)^{-1}\left({R\over D}\right)^2\!
\int_0^{2\pi}\!\!d\phi \int_0^{\pi/2}\!\!\!d\theta\,
\sin\theta\cos\theta\,
{B_\nu(\theta,\phi)\over 2},
\ee
where $(1+z)=(1-2GM/Rc^2)^{-1/2}$ is the redshift factor,
$\nu_0=\nu/(1+z)$ and $D$ is the distance to the source
(we assume the NS mass and radius are $M=1.4M_\odot$ and
$R=10$~km, respectively).

\begin{figure*}
\plotone{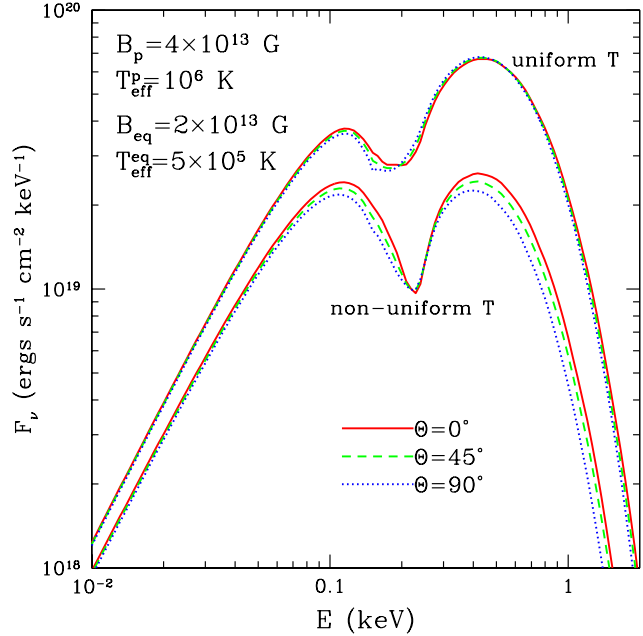}
\caption{
Synthetic spectra 
(unredshifted and with $R^2/D^2$ factored out) 
for fully ionized hydrogen atmosphere emission
from a NS with a dipole magnetic field distribution
and uniform and non-uniform [see eq.~(\ref{eq:tempdist})]
temperature distributions over the NS surface.
The uniform temperature spectrum uses the temperature profile constructed
from the atmosphere model with
$B=4\times 10^{13}$~G and $\Teff=10^6$~K,
while the non-uniform temperature spectrum uses the temperature profiles
$\Tpole$ and $\Teq$ constructed from the atmosphere models with
$B=4\times 10^{13}$~G and $\Teff=10^6$~K and
$B=2\times 10^{13}$~G and $\Teff=5\times 10^5$~K, respectively.
$\Theta$ is the angle between the observer's line of sight 
and the magnetic axis.
\label{fig:synth}
}
\end{figure*}

Figure~\ref{fig:synth} shows the computed synthetic spectra
from a fully ionized hydrogen atmosphere
with a dipole magnetic field distribution ($\Bpole=4\times 10^{13}$~G)
and uniform and non-uniform temperature distributions over the NS surface.
The proton cyclotron line at $\Ebp=0.25$~keV for
$B=4\times 10^{13}$~G is broadened significantly when the emission
is integrated over the entire NS surface due to the
variation of the magnetic field: the equivalent widths
for the two models shown are $\sim 80$ and 120~eV, as compared to
$\sim 130$ and 150~eV for models with a constant magnetic field,
and the centroid energy shifts by $\sim 10-20\%$
(roughly similar to the low $B$ results of Zane \etal~2001,
despite the differences in model and method).
The equivalent width of the line
is significant, comparable to that of the absorption features observed in 
the isolated NSs RX~J$1308.6+2127$ (Haberl \etal~2003a) and RX~J$1605.3+3249$ 
(van Kerkwijk \etal~2004) (see \S\ref{sec:discussion}).

\section{Discussion} \label{sec:discussion}

In several recent papers (Lai \& Ho~2002, 2003; Ho \& Lai~2003), we have shown
that at superstrong magnetic fields, $B\ga 10^{14}$~G, 
vacuum polarization can significantly affect the radiation
spectrum from magnetized neutron star atmospheres: it softens the 
high-energy tail of the spectrum and suppresses 
the proton cyclotron feature (above $\sim 0.6$~keV, unredshifted)
and other features (see also
Ho \etal~2003).  The latter could provide an explanation for 
the non-detection thus far of lines in the observed thermal spectra of several 
anomalous X-ray pulsars (Patel \etal~2001, 2003; Juett \etal~2002; 
Tiengo \etal~2002; Morii \etal~2003) and soft gamma-ray repeaters
(Kulkarni \etal~2003), which are thought to possess $B\ga 10^{14}$~G.

In this paper, we have studied the dependence of the neutron star atmosphere
spectrum on the magnetic field strength. As we explain qualitatively 
and show numerically, at normal neutron star magnetic fields, $B\la 10^{14}$~G, 
vacuum polarization has little effect on the atmosphere emission spectra.
Therefore, strong proton cyclotron or other atomic features may be 
present in the thermal spectrum.

Our calculations of neutron star
synthetic spectra, taking into account the line broadening effect
due to magnetic field variation over the neutron star surface, show
that the recently observed broad absorption features in several
dim isolated neutron stars
could be explained naturally as the proton 
cyclotron line (with possible blending from atomic lines of
neutral hydrogen) from neutron star atmospheres with $B\la 10^{14}$~G
(see Fig.~\ref{fig:synth}).
The variation in the strength of the observed spectral features
in these sources is then due to different fractions of the surface with
$B\la 10^{14}$~G.

For RX~J$0720.4-3125$, the weakness (phase-averaged EW $\approx 40$~eV)
of the line at 0.27~keV suggests that most of the observable surface
of this neutron star has $B\ga 10^{14}$~G,
and the line is from the magnetic equatorial region of the star
(where $B$ is weaker).
The observed line energy does not change with phase but has a
larger EW at pulse decline/minimum (Haberl \etal~2003b), which also indicates
the line is produced at a small region near the magnetic equator.
Note, however, that Kaplan \etal~(2002) and Zane \etal~(2002) place
an upper limit of $\sim\mbox{a few}\times 10^{-13}\mbox{s s$^{-1}$}$
on the period derivative of RX~J$0720.4-3125$, which, given its
period of 8.39~s, implies a dipole field with
$B_{\rm dipole}<5\times 10^{13}$~G; thus our inference of the surface
field based on the thermal spectrum implies an appreciable
non-dipolar magnetic field on the neutron star.

For RX~J$1308.6+2127$, the stronger (EW $\approx 155$~eV) line at 0.3~keV
suggests that most of the observable surface of this neutron star has
$B\la 10^{14}$~G.
This neutron star also has a double peaked pulse profile (Haberl \etal~2003a),
as compared to the very sinusoidal single peaked pulse profile of
RX~J$0720.4-3125$ (Haberl \etal~2003b).
The difference in pulse profiles may be due to differences in the viewing
geometry and angle between the magnetic and rotation axes:
in the case of RX~J$1308.6+2127$, the observer sees
the equatorial region most of the time (thus the stronger line and
slightly larger pulse fraction) and each magnetic pole,
while only one pole is visible in the case of RX~J$0720.4-3125$.

Our calculations show that, for a given line-emitting area, the 0.45~keV
line of RX~J$1605.3+3249$ should have a smaller EW than the 0.27~keV
line of RX~J$0720.4-3125$.
However, the line is stronger (EW $\approx 80$~eV) in RX~J$1605.3+3249$.
One possible way to reconcile the observed smaller EW of the 0.27~keV
line in RX~J$0720.4-3125$\footnote{de Vries \etal~(2004) find
long-term changes in the spectra of RX~J$0720.4-3125$; it is not clear
whether this variability can affect the EW of the line.}
is for the line-emitting area (with $B\la 10^{14}$~G)
to be a small fraction of the observed surface, while the
larger fraction of the surface has a much higher magnetic field
($B\ga 10^{14}$~G).
Alternatively, we have only considered hydrogen, and some of the lines
may be due to other elements, such as helium.

As discussed in \S\ref{sec:nc}, the numerical treatment of the vacuum 
polarization effect in the neutron star atmosphere models is approximate,
i.e., it does not properly account for the partial mode conversion 
associated with the vacuum resonance. Future work in this
direction is necessary (see \laihotwo), as well as a more accurate and
comprehensive study of partially ionized atmosphere models 
(see Ho \etal~2003). Furthermore, we have only examined thermal emission;
there have been cyclotron features seen in the non-thermal emission from
magnetars (e.g., the 5.0~keV feature from SGR~$1806-20$ during outburst,
Ibrahim \etal~2002, 2003; the 8.1~keV feature from AXP~1RXS~J$170849-400910$,
Rea \etal~2003).
Such theoretical/numerical studies, when combined with 
observational data, should provide useful constraints on the 
nature of various types of neutron stars, including the dim isolated neutron
stars.

\acknowledgments

We thank Alexander Potekhin for supplying the opacity tables
of Potekhin \& Chabrier (2003), which were used
in Figs.~\ref{fig:spectra} and \ref{fig:spectra3} to compute spectra
from partially ionized hydrogen 
atmospheres. We also thank Marten van Kerkwijk for useful discussion.
We appreciate the useful suggestions and comments of the anonymous
referee.
Our computations made use of the facilities at the Cornell Hewitt Computer 
Laboratory. This work is supported in part by NASA grant
NAG 5-12034 and NSF grants AST 9986740 and AST 0307252.


\end{document}